\documentclass[conference]{IEEEtran}
\IEEEoverridecommandlockouts
\usepackage{cite}
\usepackage{amsmath,amssymb,amsfonts}
\usepackage{algorithmic}
\usepackage{graphicx}
\usepackage{textcomp}
\usepackage{xcolor}
\usepackage[utf8]{inputenc}
\usepackage{subfig}
\usepackage{multirow}
\usepackage{tabularx}
\usepackage{comment}
\usepackage{amsmath}
\def\BibTeX{{\rm B\kern-.05em{\sc i\kern-.025em b}\kern-.08em
    T\kern-.1667em\lower.7ex\hbox{E}\kern-.125emX}}
\newcolumntype{L}[1]{>{\raggedright\let\newline\\\arraybackslash\hspace{0pt}}m{#1}}
\newcolumntype{C}[1]{>{\centering\let\newline\\\arraybackslash\hspace{0pt}}m{#1}}
\newcolumntype{R}[1]{>{\raggedleft\let\newline\\\arraybackslash\hspace{0pt}}m{#1}}

\begin{document}

\title{Going Deep: Using deep learning techniques with simplified mathematical models against XOR BR and TBR PUFs (Attacks and Countermeasures)}
\author{\IEEEauthorblockN{Mahmoud Khalafalla, Mahmoud A. Elmohr, Catherine Gebotys}
\IEEEauthorblockA{Department of Electrical and Computer Engineering\\
University of Waterloo\\
Ontario, Canada\\
Email: \{mahmoud.khalafalla, mahmoud.elmohr, cgebotys\}@uwaterloo.ca}

}

\maketitle

\begin{abstract}
 This paper contributes to the study of PUFs vulnerability against modeling attacks by evaluating the security of XOR BR PUFs, XOR TBR PUFs, and obfuscated architectures of XOR BR PUF using a simplified mathematical model and deep learning (DL) techniques. DL modeling attacks were invoked against PUFs with different stage sizes (e.g. 64, 128, 256) and all are implemented on FPGA chips. Obtained results show that DL modeling attacks could easily break the security of 4-input XOR BR PUFs and 4-input XOR TBR PUFs with modeling accuracy $\sim$ 99\%. Similar attacks were executed using single-layer neural networks (NN) and support vector machines (SVM) with polynomial kernel and the obtained results showed that single NNs failed to break the PUF security. Furthermore, SVM results confirmed the same modeling accuracy reported in previous research ($\sim \textbf{50\%}$). For the first time, this research empirically shows that DL networks can be used as powerful modeling techniques against these complex PUF architectures for which previous conventional machine learning techniques had failed. Furthermore, a detailed scalability analysis is conducted on the DL networks with respect to PUFs' stage size and complexity. The analysis shows that the number of layers and hidden neurons inside every layer has a linear relationship with PUFs' stage size, which agrees with the theoretical findings in deep learning. Consequently, A new obfuscated architecture is introduced as a first step to counter DL modeling attacks and it showed significant resistance against such attacks (16\% - 40\% less accuracy). This research provides an important step towards prioritizing the efforts to introduce new PUF architectures that are more secure and invulnerable to modeling attacks. Moreover, it triggers future discussions on the removal of influential bits and the level of obfuscation needed to confirm that a specific PUF architecture is resistant against powerful DL modeling attacks.
\end{abstract}


\section{Introduction}
For more than 15 years, physically unclonable functions (PUFs) have been considered as promising cryptographic primitives that can be used in a wide range of security applications, such as identification, authentication, and cryptographic key generation\cite{Lee_2004}\cite{Lim_2004}\cite{Maes_2015}.

Intrinsic electronic PUFs are the most widely proposed architectures in literature because they are more secure and easier to realize on silicon chips without equipment and processing overhead\cite{Bohm_2012}. There exist two main types of intrinsic PUFs, where they both depend on delay measurements by either using an arbitration circuit (delay-based PUFs) or the bi-stability property of memory cells (memory-based PUFs). Delay-based strong PUFs depend on symmetrical paths where a signal traverse these paths depending on an input challenge and the response depends on an arbitration circuitry that determines which path is shorter. The advantage of this PUFs type is the large input challenge-response space which makes it hard to use brute force attacks to break their security. However, delay-based PUFs have response reliability issues when operating under temperature and voltage variation conditions \cite{Katzenbeisser_2012}. Moreover, most of the architectures showed vulnerability against modeling attacks using their collected challenge-response pairs (CRPs) for training\cite{Ruhrmair_2010}\cite{Ruhrmair_2013}. 

Proposals of new PUF architectures tried to merge between delay-based and memory-based approaches to design a stronger PUF with more resistance against modeling attacks and large challenge space, so CRPs cannot be exhaustively read by the attackers.  One of these proposals is the Bistable Ring PUF (BR PUF), introduced in \cite{BR_intro} and \cite{Chen_2012}. Its basic idea is that the output of any given ring with an even number of inverters has only two possible stable states. This is similar to memory-based PUFs operation except that challenge bits are inserted to select which path to be used at every stage (more details in the next section). One problem of BR PUFs is that it takes a longer time to stabilize, which is an undesirable property of PUFs. Furthermore, BR PUFs implementations on FPGAs showed an output bias problem as reported in \cite{TBR_intro}. As a result, other variations of BR PUFs were proposed like the Twisted Bistable Ring PUFs (TBR PUF) \cite{TBR_intro} and the XOR BR PUFs in \cite{Xu_2015}. The latter was proposed after successful modeling attacks were reported against BR PUFs and TBR PUFs using SVM and single-layer Artificial Neural Network (ANN). XOR BR PUFs showed significant resistance against modeling attacks and set an example that the approach of complicating the relationship between input challenges and responses can somehow countermeasure conventional ML modeling techniques used to break the security of previous architectures. However, it was shown in \cite{Ganjietal_2016} that BR PUF families have a finite set of influential challenge bits and can be considered a Linear Threshold Function (LTF) similar to Arbiter PUFs (APUFs). Hence, an XOR version of BR PUFs can be modeled using a single-layer perceptron function as was reported in \cite{Ganji_2016_2} against XOR APUFs. In this paper, we show that XOR BR and XOR TBR PUFs cannot be modeled using a single layer NN and we needed to use deeper architecture along with a simplified mathematical model to fully break their security.

Our motivation is based on the current status of PUFs design efforts, which part of the focus on introducing new architectures solely depending on more complex challenge-response relationships, and measure the architectural enhancement by how resistant these PUFs are to conventional ML modeling attacks. However, the development in DL techniques and the hardware running them with acceptable timing performance put more pressure and challenges on those PUF architectures.
Hence, the contributions of this paper can be listed as follows:
\begin{enumerate}
	\item Showing that XOR BR and XOR TBR PUFs cannot be modeled using single NNs and justifying the necessity to use deeper networks.  
	\item Applying deep learning modeling techniques to attack 4-input XOR BR PUF, XOR TBR PUF, and an obfuscated version of XOR BR PUF and successfully breaking their security with different stage sizes (e.g. 64, 128, 256) and achieving a remarkable modeling accuracy $>$ 99\%.
	\item Taking the first step to connect between DL theory and our empirical results to understand why DL attacks work successfully and based on that we introduced a new obfuscated version of XOR BR PUF as a countermeasure.
	\item Providing detailed analysis on the scalability of DL modeling techniques in terms of layers number and hidden neurons needed with respect to the PUF architecture complexity and number of PUF stages.  
	
\end{enumerate}

The rest of the paper is organized as follows. Section~\ref{Background} discusses the background and further motivation for our work. Sections~\ref{sec:Hardware} and ~\ref{sec:obfuscated} show the detailed hardware implementation of XOR BR PUFs, XOR TBR PUFs and the obfuscated versions of XOR BR PUF. Section~\ref{DNN-Arch} describes in detail the DL network architecture, experiments setup and all decisions taken to apply a successful modeling attack. Obtained results and discussions on DL and attack practicality are provided in Section~\ref{results} followed by the conclusion in Section~\ref{conclusion}. 

\section{Background and Motivation}\label{Background}

\subsection{Bistable ring PUFs}\label{BRPUF}
\begin{figure}[ht]
	\centering
	\includegraphics[trim=0.5cm 0cm 0cm 1.3cm, clip=true,width=0.8\linewidth]{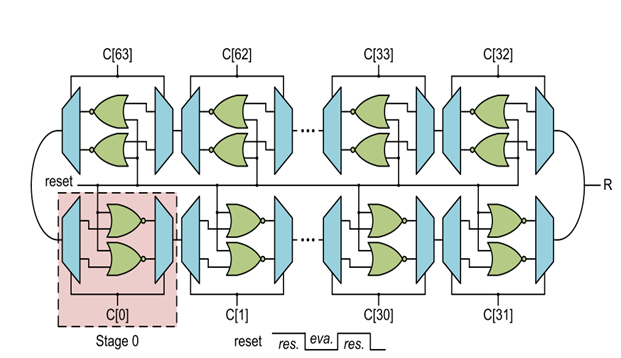}
	\caption{BR PUF Architecture \cite{BR_intro}}
	\label{fig:fig02}
\end{figure}

Numerical Modeling attacks against delay-based PUFs were one of the earliest attacking techniques proposed in the literature\cite{Ruhrmair_2010}. Given a set of CRPs and using machine learning techniques like SVM, Logistic Regression (LR), and Evolution Strategy (ES), it is possible to accurately predict the PUF outcome for the whole challenge-response space. In \cite{Ruhrmair_2010} and \cite{Ruhrmair_2013}, Ruhrmair U. et al introduced an accurate mathematical model for APUFs that can be exploited by machine learning techniques to model PUFs response. This parametric model could also be expanded for other delay-based PUFs (e.g XOR PUFs, FeedForward PUFs). Therefore, the security of most delay-based PUF architectures could be broken \cite{Ruhrmair_2010},\cite{Hospodar_2012},\cite{Ruhrmair_2013}, and\cite{Tobisch_2015}. Hence, BR PUF was introduced as a new hybrid architecture to countermeasure the modeling attacks on delay-based PUFs. The new idea was to place an even number of PUF stages in a closed loop and the input challenge bits will determine which gates will be involved in every stage. As shown in Figure~\ref{fig:fig02}, every stage has two NOR gates working as inverters, a multiplexer and a de-multiplexer controlled by the challenge bits to determine which inverters contribute to the loop. The sequence of operations starts by setting the reset signal to '1' to force all stages to output '0', then applying the challenge bits, and clearing the reset signal to '0', and finally waiting for the ring output to be stable before reading it out. 

Since the BR PUF architecture is operating like memory-based PUFs, it was expected that no mathematical model could be built for such architecture. However, there exist several reported machine learning attacks against BR PUFs by either using a simplified mathematical model \cite{Xu_2015}\cite{TBR_intro} or without a model \cite{Ganjietal_2016}. Furthermore, it was found that the BR PUF's responses were not uniform and are biased \cite{Xu_2015}. Hence, the TBR PUF was introduced by Schuster D. and Hesselbarth R. after successfully breaking the security of the BR PUF using a single layer NN and a simplified mathematical model (i.e. replacing every '0' challenge bit with '-1') \cite{TBR_intro}. This new variant of BR PUF makes all inverters involved in the closed-loop, and the challenge bits are responsible for determining the positions of inverters inside the ring (odd or even). 

Although this architectural modification showed more resistance against ANN attacks and more uniform responses, in the same paper, it was noted that NNs were learning the correlation between challenge bits and responses and there might exist more influential bits that helped in modeling BR and TBR PUFs. Gangi F. et al \cite{Ganjietal_2016} proposed a new attack against BR and TBR PUFs that does not require deriving a mathematical model of the BR PUF family. The main idea is to exploit the challenge bits with higher influence on the PUF responses (influential challenge bits) to construct a machine-learning-based boosted model that can predict the PUF outcome with high probability. The experiments were conducted using 30K CRPs collected from 64 stages BR and TBR PUF implementations on Altera Cyclone IV FPGAs, and the adaptive boost algorithm \cite{Freund_1997} was used to create the boosted classifier built over the initial weak learners, which depends on single influential bits. Obtained results showed that the boosting technique could successfully model both PUFs up to 99\% prediction accuracy using 50 boosting iterations. Furthermore, they suggested that BR and TBR PUFs have a small set of influential bits and their polynomial threshold function (PTF) can be approximated by an LTF. 

\subsection{XOR Bistable ring PUFs}\label{XOR_BRPUF}
 In \cite{Xu_2015}, Xu X. et al could build a simplified mathematical model to attack BR and TBR PUFs using SVM modeling technique. Instead of deriving an accurate non-linear model of PUF delays, A simplified additive model was adopted to represent the difference between the pull-up and pull-down strength of every inverter (represented by a NOR gate in Figure~\ref{fig:fig02}). Hence, the PUF response can be re-written as the summation of all stages' strength difference and therefore, SVM works by learning the weights assigned to every inverter strength difference. Equations~\ref{equation:BR_1} to ~\ref{equation:BR_3} explain the model parameters, where '$t_{i}$' and '$b_{i}$' are the pull-up and pull-down strength difference for the top and bottom NOR gate at the $i^{th}$ stage. Hence, an even stage will contribute to a positive PUF response with strength $ t_{i}$ or $b_{i} $ depending on the challenge bit value and odd stages will contribute with strength $-t_{i}$ or $-b_{i}$.  A generalization of these terms can be used to represent the odd and even stages contribution in this form '$-1^{i}t_{i}$' and '$-1^{i}b_{i}$'. In equation~\ref{equation:BR_2}, Xu X. et al defined two terms '$\alpha_{i}$' and '$\beta_{i}$' to facilitate the writing of PUF response summation equation with respect to input challenge bits. Hence, PUF response can be represented as a linear summation as shown in equation~\ref{equation:BR_3}, where 'K' is the number of PUF stages and '$C_{i}$' $\in {-1,1}$ is the challenge bit at this stage (note that 0 value is interpreted as -1 to select the bottom NOR gate). Furthermore, the term '$\alpha_{i}$' can be discarded because it yields the same value for all CRPs training samples.
 
\begin{equation}
\label{equation:BR_1}
\Delta Strength_{upper}^{i} = -1^{i}t_{i},\enskip \Delta Strength_{buttom}^{i} = -1^{i}b_{i}
\end{equation}

\begin{equation}
\label{equation:BR_2}
\begin{split}
\alpha_{i} + \beta_{i} = -1^{i}t_{i} \quad,\quad \alpha_{i} - \beta_{i} = -1^{i}b_{i} \textrm{ , hence } \\ \alpha_{i} = -1^{i}(\dfrac{t_{i} + b_{i}}{2}) \quad,\quad \beta_{i} = -1^{i}(\dfrac{t_{i} - b_{i}}{2})
\end{split}
\end{equation}

\begin{equation}
\label{equation:BR_3}
T_{response} = Sign(\sum_{i=0}^{K} (\alpha_{i} + C_{i}\beta_{i}))
\end{equation}
\\
The experimental results showed that BR and TBR PUFs could be successfully attacked using SVM with modeling accuracy $>95\%$. Hence, XOR BR PUF was introduced to countermeasure this attack and the results showed that the SVM modeling technique with polynomial kernel was not successful in breaking architectures with XOR input $>$ 3 as shown in Table~\ref{tab:XOR_BR_result_xu}.

\begin{table}[ht]
	\caption{Reported results of SVM modeling attack on XOR BR PUF \cite{Xu_2015}.}
	\label{tab:XOR_BR_result_xu}
	\begin{tabular}{c|c|c|c}
		\hline 
		 XOR inputs & Stages & Training Size & Modeling Accuracy \\ 
		\hline 
		3 & 32 & 1200 & $>$ 95\% \\ 
		\hline 
		3 & 64 & 7200 & $>$ 95\% \\ 
		\hline 
		3 & 128, 256 & N/A & 50.1\% \\ 
		\hline 
		4 & 32,64,128,256 & N/A & 50.1\% \\ 
		\hline 
	\end{tabular} 
\end{table} 

\subsection{Motivation for deep learning attacks}\label{Motive}
\begin{figure}[ht]
	\centering
	\includegraphics[trim=2.5cm 2cm 2.5cm 1.8cm, clip=true,width=0.8\linewidth]{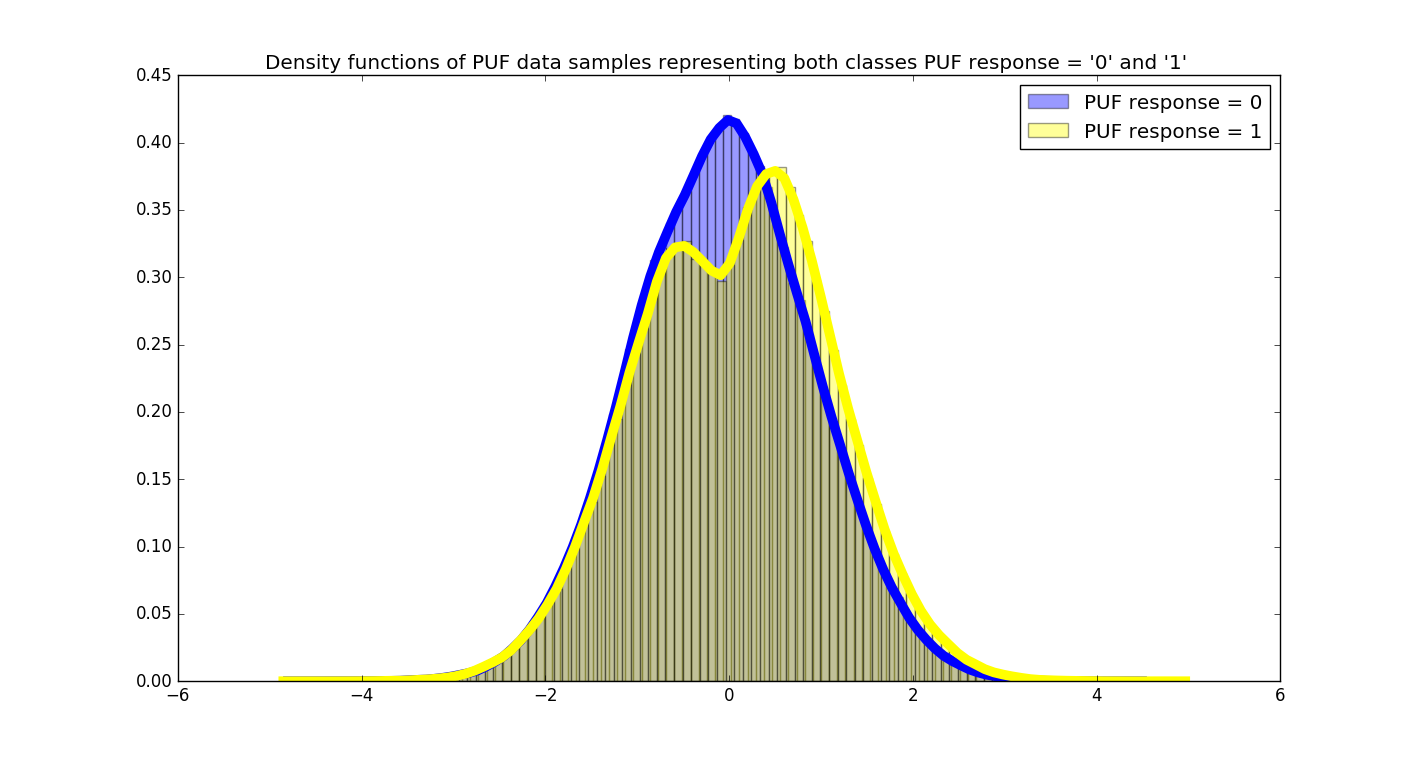}
	\caption{Density function of LDA feature analysis for 128 stage 4-input XOR BR PUF}
	\label{fig:fig04}
\end{figure}

Gangi F. et al \cite{Ganji_2016_2} showed that XORed LTFs (e.g. APUFs) with the number of XOR inputs $<$ ln(number of PUF stages) can be learned using single-layer perceptron function in polynomial time. Consequently, if BR PUFs can be approximated by an LTF because of the finite set of influential bits as discussed earlier, then it is expected that using a single layer NN can model XOR BR PUFs. However, we conducted an analysis on the implemented XOR BR and TBR PUF instances using linear discriminant analysis (LDA) to confirm if both classes representing PUF response are linearly separable or not. LDA is a supervised linear transformation technique used to reduce features dimensionality by computing the linear discriminants or the directions of the axes at which, the separation between multiple classes is maximized \cite{hastie_01}. As a result, all XOR BR and TBR PUF instances with different sizes (64, 128, 256) showed similar behavior to the example in Figure~\ref{fig:fig04}. It shows the density function of 1M data samples representing both PUF response classes '0' and '1'. It is clear that both curves are overlapping and hence, they are not linearly separable, and a single layer perceptron algorithm cannot model this type of architectures. One might attribute the different behavior of XOR BR PUFs compared to XOR APUFs to the fact that we could derive an accurate additive linear model for the latter. On the other hand, the mathematical model of XOR BR PUFs is a simplified one. It is shifting from modeling delay difference to a more abstract concept of modeling the strength difference of every stage as discussed earlier. Furthermore, this also justifies why SVM with polynomial kernel could not break XOR BR PUF security while it was reported in previous literature that similar XOR LTFs (i.e. APUFs) could be broken using logistic regression technique (LR)\cite{Ruhrmair_2010}\cite{Ruhrmair_2013}.

Hence, the motivation behind this work was to explore the ability of deep learning modeling techniques to break the security of XOR BR PUFs. Furthermore, experiments were executed to attack XOR TBR PUFs since they showed more resistance to ANN modeling attacks compared to BR PUFs. Therefore, XOR TBR PUF was expected to be harder to break and introduces an extra challenge.

\section{Hardware Implementation on FPGA}
\label{sec:Hardware}

\subsection{Overall system architecture}
\label{sec:architecture}

\begin{figure}[ht]
	\centering
	\includegraphics[trim=2.5cm 3.5cm 2.5cm 3.5cm, clip=true, width=0.8\linewidth,height=2.8cm]{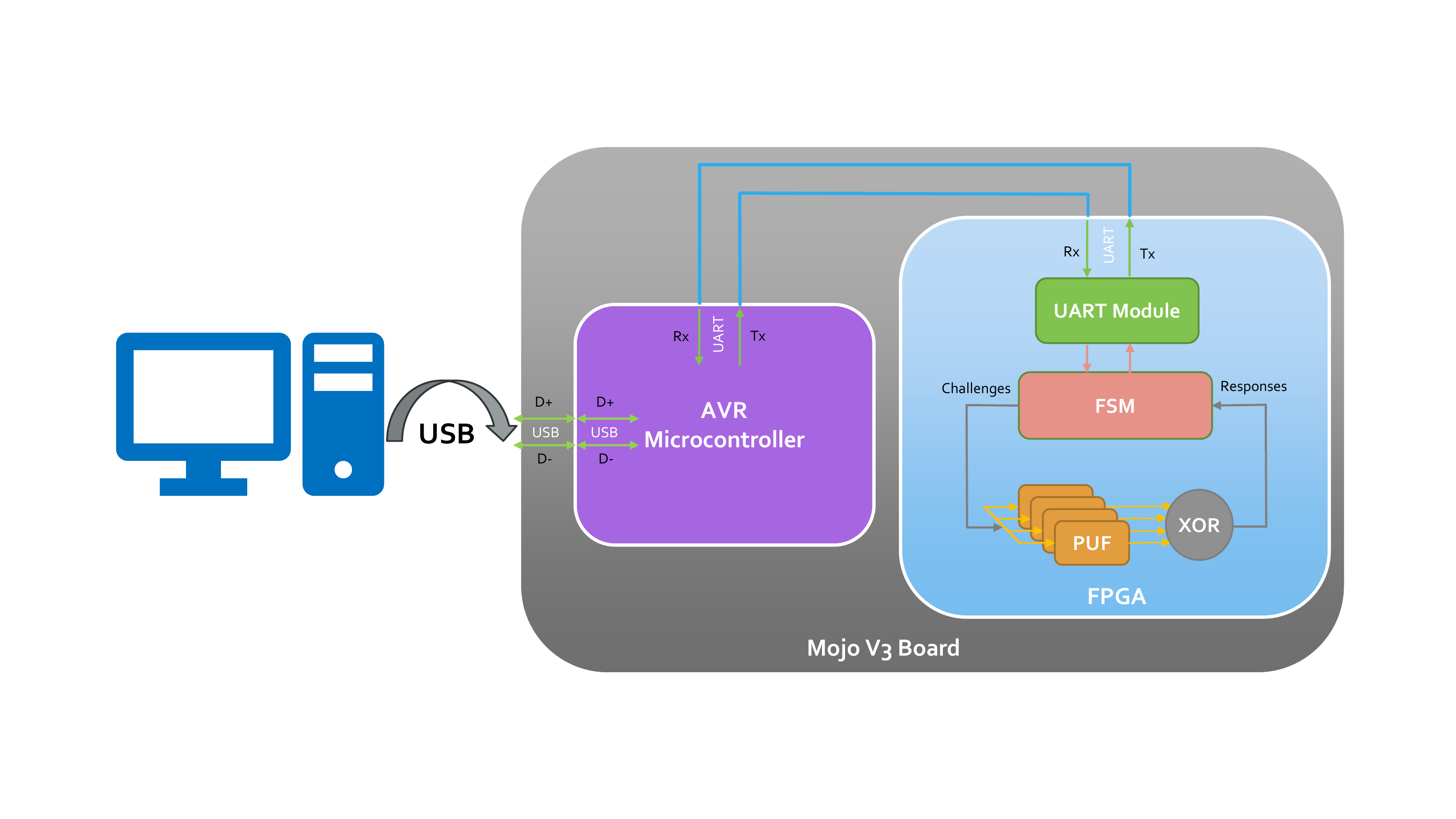}
	\caption{PUF Ecosystem} 
	\label{fig:ecosys_fig} 
\end{figure}

In our implementation, we used the Mojo V3 Board which features a Spartan-6 Xilinx FPGA alongside an AVR microcontroller. As shown in Figure~\ref{fig:ecosys_fig}, the PUF is implemented on the FPGA side in addition to a Finite State Machine (FSM) for control as well as a UART module for receiving challenges and sending responses. The UART module on top of the FPGA is connected to the UART of the AVR microcontroller, which in return transfers the data through the USB port of the Mojo board to and from an external PC.

A script employed on the PC side is responsible for generating and sending random challenges to the board, receiving the responses and storing them into log files in order to evaluate the PUF’s characteristics as will be discussed in Section~\ref{sec:characteristics}. 

Challenges are generated using a Galois Linear-Feedback Shift Register (LFSR) pseudo-random number generator on the PC side and sent to the PUF through serial communication to the FPGA byte by byte, where the FSM stores each byte and concatenates them to form the full challenge vector. After the FSM forms the full challenge vector, it transfers it to the PUF at once as well as releasing the reset signal for the PUF at the same time. Then the FSM waits for a predetermined time so that the PUF converges to a stable state before capturing the response. BR and TBR PUFs’ response is usually the output of one stage of the ring, however, in our implementation, we derived all the stages’ outputs to make sure that all stages have the same capacitive load, as otherwise one stage would have different load than others, which might bias the PUF. The same note was considered in previous literature in~\cite{capacitance}. Another advantage of deriving all stages’ outputs is to distinguish between converged and non-converged responses. To do so, the response bytes are ORed together forming one byte that is sent to the PC side. For a converged ring, that byte should be either ‘10101010’ (0xAA) representing '1' or ‘01010101’ (0x55) representing '0', other than that, it indicates a non-converged ring, which should not be added to the CRPs database.

\subsection{Implementation approaches}
\label{sec:implementation}

\begin{figure}[ht]
	\centering
	\subfloat[]{
		\label{fig:BR_layout}
		\includegraphics[trim=9cm 4cm 9cm 4cm, clip=true, width=0.38\linewidth]{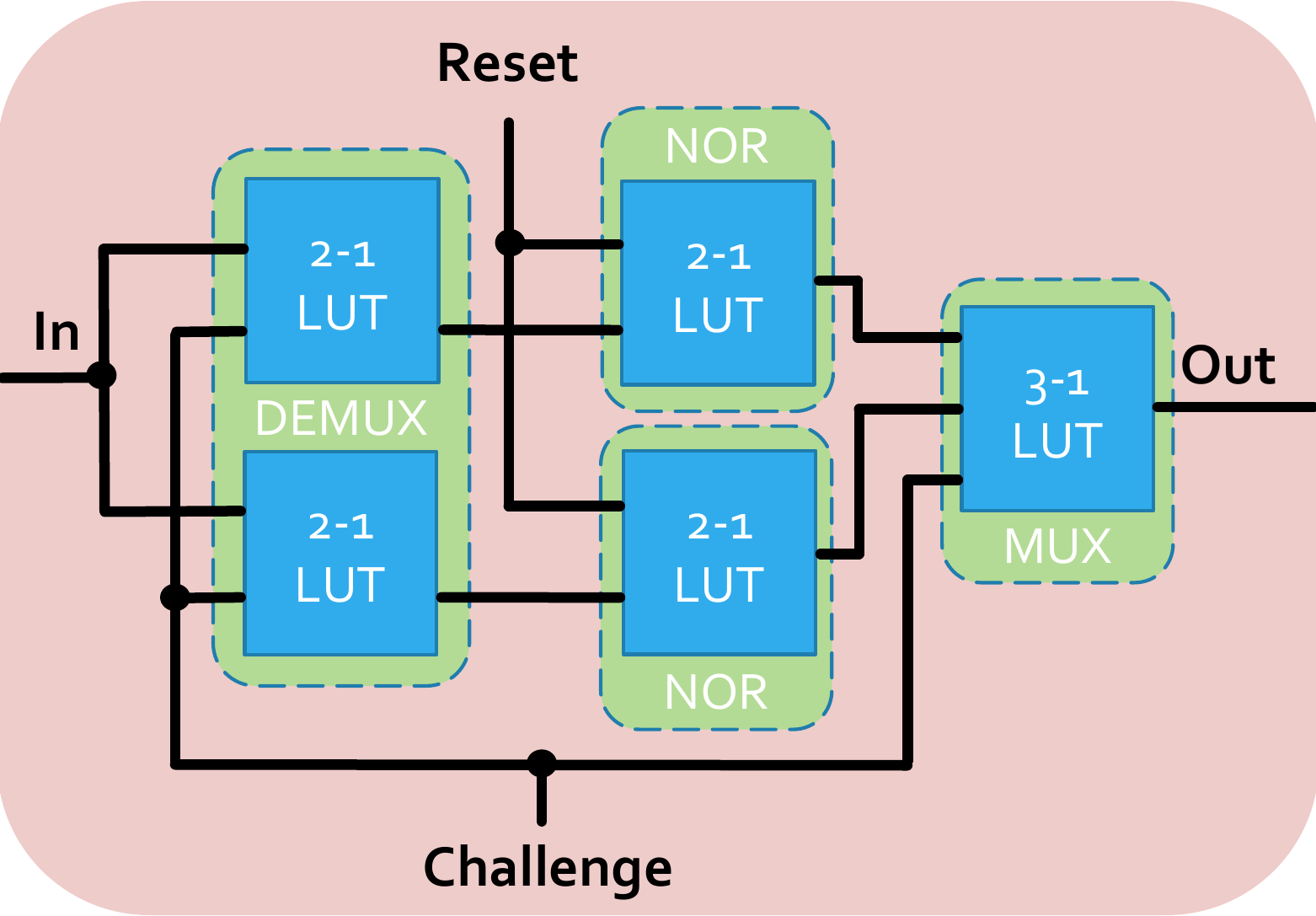}
	}
	\hspace{1mm}
	\subfloat[]{
		\label{fig:BR_opt_layout}
		\includegraphics[trim=10.9cm 4.3cm 10.9cm 4.3cm, clip=true, width=0.305\linewidth]{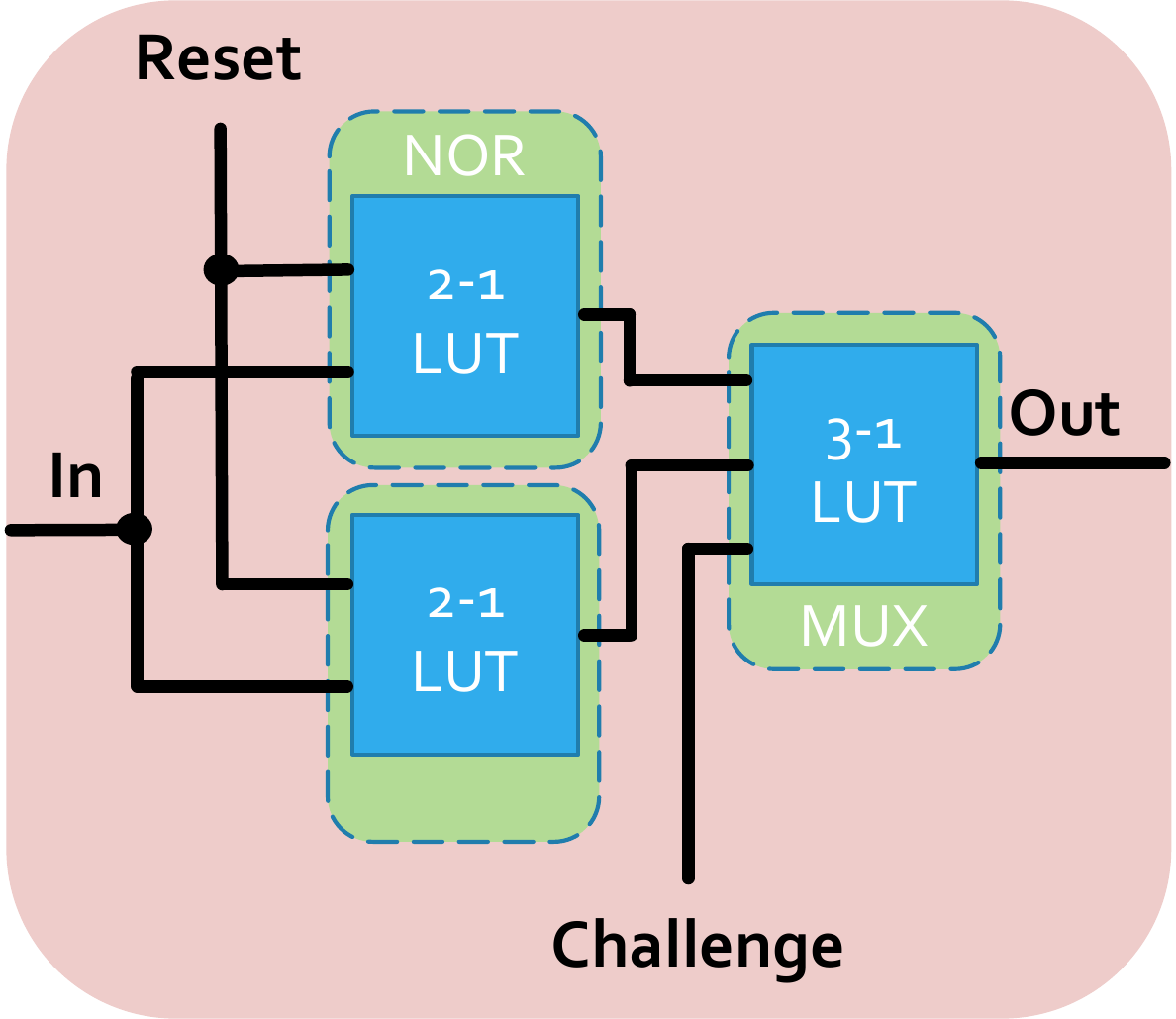}
	}
	\\
	\caption{BR PUF Optmization Layout on FPGA}
	\label{BR_Layout} 
\end{figure}

A single stage of the BR PUF as introduced in~\cite{BR_intro} should contain one MUX, one DEMUX and two inverting elements such as NOR gates as in Figure~\ref{fig:fig02}. A straight forward implementation on the FPGA would result in five Look Up Tables (LUTs) as in Figure~\ref{fig:BR_layout} which was the approach used in~\cite{capacitance}. In our implementation though, we introduced an optimization to the architecture level by removing the DEMUX. As eventually the MUX chooses between the output of either NOR gates, which is the important issue for the BR PUF functionality, however, whether supplying the input to only one NOR gate as in the original design or to both NOR gates as in our optimized design, it would not make a difference to the BR PUF functionality. With this optimization the number of LUTs is reduced to three as in Figure~\ref{fig:BR_opt_layout}.

\begin{figure}[ht]
	\centering
	\subfloat[]{
		\label{fig:TBR_layout}
		\includegraphics[trim=7.4cm 4cm 7.4cm 4cm, clip=true, width=0.46\linewidth]{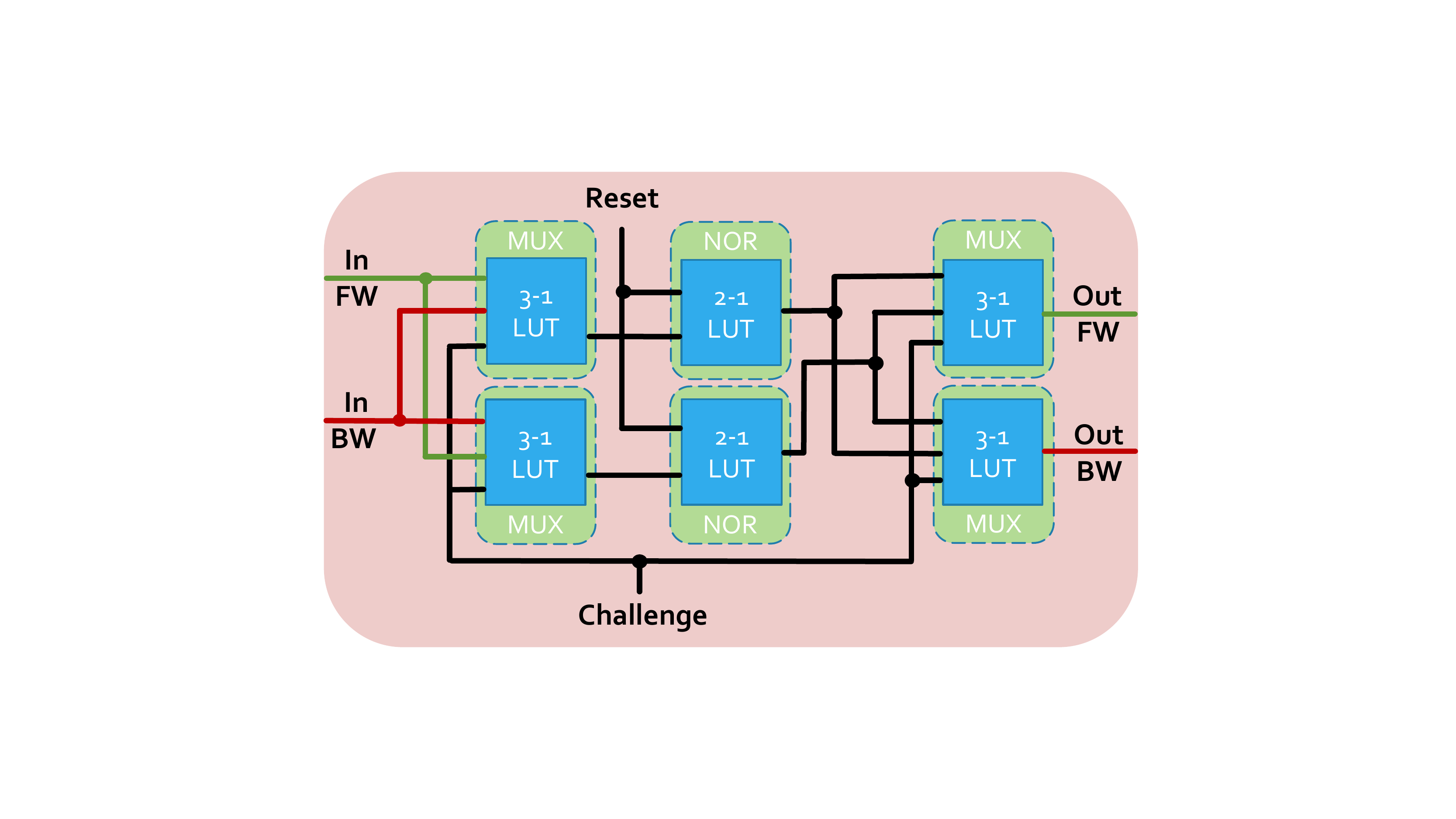}
	}
	\hspace{1mm}
	\subfloat[]{
		\label{fig:TBR_opt_layout}
		\includegraphics[trim=9cm 3.9cm 9cm 3.9cm, clip=true, width=0.38\linewidth]{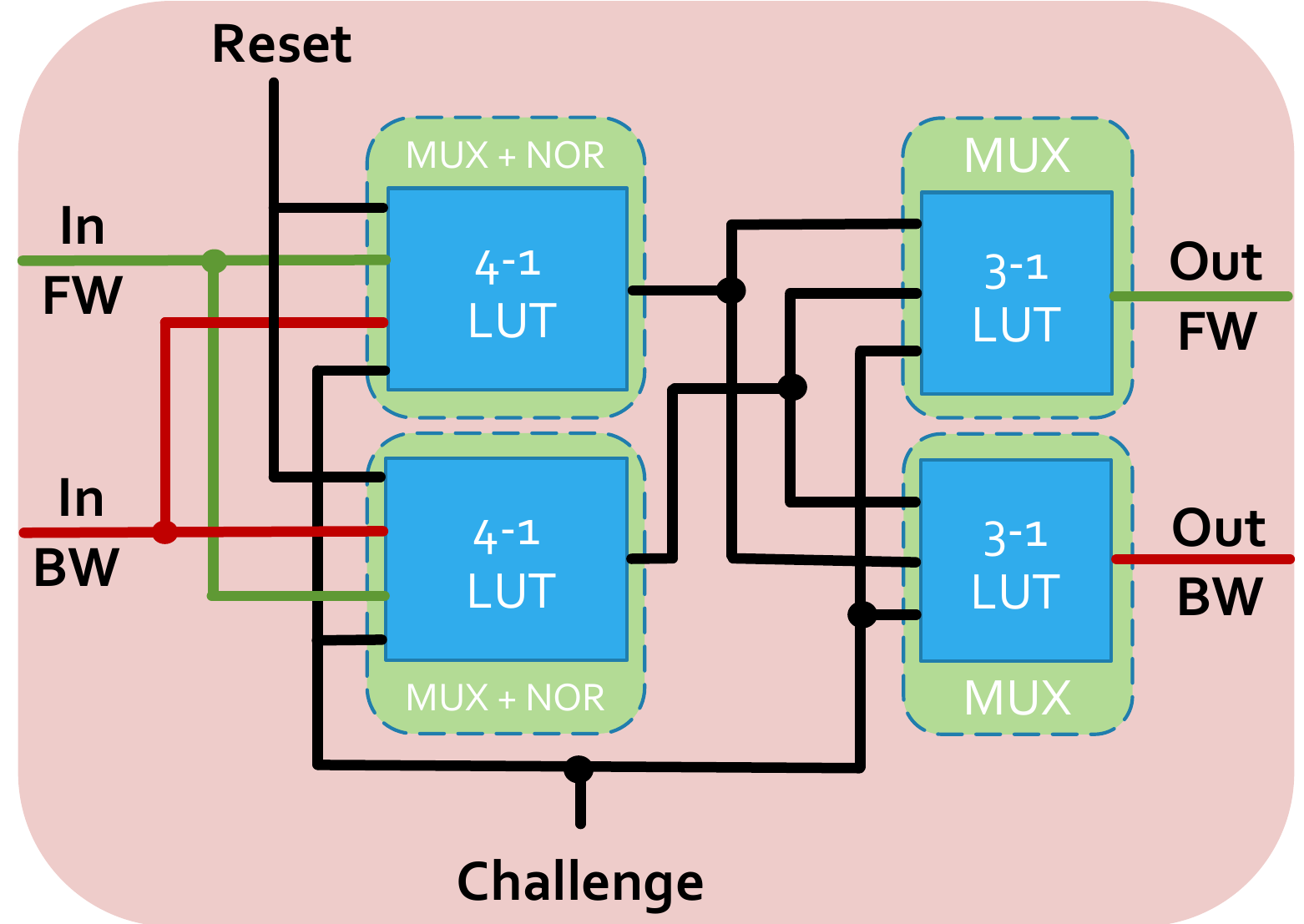}
	}
	\\
	\caption{TBR PUF Optmization Layout on FPGA}
	\label{TBR_Layout} 
\end{figure}

For the TBR PUF introduced in~\cite{TBR_intro}, the components of a single stage can be interpreted as 2 MUXes feeding the 2 NOR gates and two other MUXes choosing between the outputs of the two NOR gates which was also adopted in~\cite{Ganjietal_2016}. A straight forward implementation would try to transform each component of the PUF into a separate LUT as in Figure~\ref{fig:TBR_layout} resulting in a total of 6 LUTs for one TBR PUF stage. However, in our implementation, we merged each NOR gate with its preceding MUX as shown in Figure~\ref{fig:TBR_opt_layout} resulting in a total of only 4 LUTs for one TBR PUF stage. This optimization does not affect the functionality of the TBR PUF as it maintains two inverting elements and two different paths, in addition to the fact that the merged LUT actually has the same logic of the two separate LUTs combined.

The reasons behind these optimizations were that both PUFs with 128-bit and 256-bit challenge were not possible to be implemented on the FPGA without the optimization due to the limited LUTs on the FPGA. Preliminary experiments were conducted over the non-optimized implementations for 64-bit BR and TBR PUFs to empirically confirm that the optimizations presented for both BR and TBR PUFs do not affect the security of the PUFs. The obtained results showed that non-optimized architectures showed similar performance against both SVM and DL techniques. Thus, these optimizations do not change the PUFs’ architectures and empirically do not affect their security.

\subsection{PUF characteristics}
\label{sec:characteristics}

A good PUF should achieve three main properties: reliability, unpredictability, and uniqueness~\cite{metrics}. In this section, we define the evaluation metrics that measure these three properties and provide the actual characteristics for the implemented PUFs. These metrics are PUF noise, PUF bias, and inter-chip hamming distance. We also considered another important characteristic which is individual challenge bits influence.

\begin{enumerate}
	\item PUF Noise:
	A reliable PUF would give a consistent response to a certain challenge per chip, however, in reality, a PUF might give inconsistent responses for the same challenge. 
	To measure noise, we apply the same challenge to the same chip for many iterations, take a majority vote to determine the supposedly right response and repeat that for all challenges. Thus we can calculate the noise as in equation~\ref{equation:noise} with an ideal value of 0.
	
	\begin{equation}
    \label{equation:noise}
    N = \frac{\sum \#\ wrong\ responses}{\#\ iterations \times \#\ challenges}
    \end{equation}
    \hspace{1cm}

	\item PUF Bias:
	Which represents the tendency of the PUF to respond with 0 or 1 to different challenges. Bias can be calculated as in equation~\ref{equation:bias} with an ideal value of 0.5.  
	
	\begin{equation}
    \label{equation:bias}
    B = \frac{\#\ responses\ of\ '1'}{\#\ challenges}
    \end{equation}
    \hspace{1cm}

	\item Inter-Chip Hamming Distance:
	Different chips should give different responses for the same challenge. Inter-chip hamming distance represents how many responses were dissimilar for the same challenge on different chips. The normalized hamming distance between two different chips would be calculated as in equation~\ref{equation:NHD} with an ideal value of 0.5.  
	
	\begin{equation}
    \label{equation:NHD}
    NHD = \frac{\#\ dissimilar\ responses}{\#\ challenges}
    \end{equation}
    \hspace{1cm}
    
	\item Individual Challenge Bits Influence:
	Challenge bits should ideally contribute equally to the resulted response, not only some of the challenge bits. For each challenge bit, its influence is calculated as in equations~\ref{equation:inf0} and~\ref{equation:inf1} with an ideal value of 0.5 for both.  
	
	\begin{equation}
    \label{equation:inf0}
    Infl(i,0) = \frac{\#\ responses\ of\ '1'}{\#\ challenges\ with\ ith\ bit = 0}
    \end{equation}

    \begin{equation}
    \label{equation:inf1}
    Infl(i,1) = \frac{\#\ responses\ of\ '1'}{\#\ challenges\ with\ ith bit = 1}
    \end{equation}
    \hspace{1cm}
\end{enumerate}

To obtain the actual characteristics of the implemented PUFs detailed in Table~\ref{tab:results}, we used three typical Mojo boards, loaded the same PUF design on all of them and applied 1 Million different challenges each for three iterations. All experiments were conducted in room temperature. 

An important note is that as we had the ability to distinguish between converged and non-converged responses, we excluded the non-converged responses from characteristics calculations. Moreover, training and testing sets contained only the converged responses after the majority vote, thus eliminating PUFs noise for the neural networks. 
It is important to note that all results presented in Table~\ref{tab:results} are for the XORed PUFs treated as a black box, not for individual PUFs. Also, the reported characteristics are averages over the three chips, except for the bias being reported for each chip.

\begin{table}[ht]
	\caption{Implemented PUFs Characteristics}
	\label{tab:results}
	\begin{center}
		\begin{tabular}{ c | c | c | c | c | c | c }
			\hline
			\multirow{3}{*}{Characteristic} & \multicolumn{6}{ c }{PUF Size \& Type} \\ \cline{2-7}
			& \multicolumn{3}{ c |}{BR PUF} & \multicolumn{3}{ c  }{TBR PUF} \\ 
			
			& \multicolumn{1}{c}{64} & \multicolumn{1}{c}{128} & \multicolumn{1}{c|}{256} & \multicolumn{1}{c}{64} & \multicolumn{1}{c}{128} & \multicolumn{1}{c}{256}  \\ \hline
			
			Eval. Time (Cyc.)  & 1024 & 4000 & 9600 & 6000 & 13400 & 19000 \\ \hline
			Conv. Avg. (\%) & 81 & 82 & 86 & 74 & 73 & 69 \\ \hline
			Noise Avg.  (\%) & 1 & 2 & 1 & 3 & 3 & 3 \\ \hline
			Bias Chip 1 (\%) & 48 & 50 & 47 & 51 & 47 & 53 \\ \hline
			Bias Chip 2 (\%) & 49 & 49 & 54 & 54 & 52 & 47 \\ \hline
			Bias Chip 3 (\%) & 47 & 48 & 53 & 53 & 53 & 49 \\ \hline
			NHD Avg. (\%) & 49 & 53 & 52 & 50 & 47 & 52 \\ \hline		
			Max Infl.  (\%) & 43 & 47 & 58 & 59 & 56 & 45 \\ \hline

			\hline
		\end{tabular}
	\end{center}
\end{table}

As shown in Table~\ref{tab:results}, the implemented PUFs have near-ideal characteristics for bias and inter-chip hamming distance. Even noise is negligible because we eliminated the non-converged responses. The non-converged responses ranged from 14\% to 31\% corresponding to convergence rates between 86\% and 69\%, which is one of the drawbacks of BR and TBR PUFs. The reported cycles spent waiting for convergence before capturing responses (the evaluation time) are the minimum cycles needed to achieve the corresponding convergence rates, (waiting for more time would not lead to any significant improvement in the convergence rates).

Also as shown there are no individual influential bits. The maximum influence a challenge bit can get is 59\% which is not a huge influence compared to the ideal influence of 50\%.

\section{Obfuscated XOR BR PUF Architectures as Countermeasures Against DL Attacks}\label{sec:obfuscated}
This section provides an overview of two obfuscated XOR BR PUF architectures implemented to show how DL modeling attacks perform against challenge obfuscation techniques and how to develop countermeasures resistant to such attacks. the obfuscation logic elements including memory-based PUFs and multiplexers in both architectures are implemented in software as a proof of concept. Hence, the original challenge is supplied as input to the obfuscation logic and the output modified challenge is sent to the hardware XOR BR PUFs as mentioned in Section~\ref{sec:architecture}. We adopted this approach because it is easier to implement and reuse the deployed XOR BR PUF instances with no impact on hardware functionality. Furthermore, We are interested in the logical reasoning of how to design a new architecture to thwart DL modeling attacks. Further hardware and power overhead analysis is beyond the paper scope.

\subsection{Obfuscated PUF architecture 1 (Hierarchical XOR BR PUF)}

\begin{figure}[ht]
	\centering
	\includegraphics[trim=0cm 0cm 0cm 0cm, clip=true,width=0.8\linewidth]{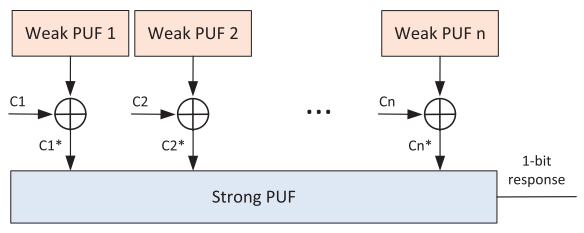}
	\caption{The Multi-PUF Architecture introduced in \cite{Ma_2018}}
	\label{fig:fig_MPUF}
\end{figure}

This architecture adopts a similar obfuscation technique to the one introduced in \cite{Ma_2018} by using a pool of memory-based PUFs responses to XOR with the original challenge. Note that memory-based PUFs should have a 50\% randomness (50\% responses are '1'), therefore, nearly half of the challenge bits will be inverted and the rest will pass through without change. The XOR output result will be the new challenge to the XOR BR PUF as shown in Fig~\ref{fig:fig_MPUF}. The PUF statistical properties are similar to those in Table~\ref{tab:results} because the same PUF instances were used. Although this technique hides the input PUF challenge, it suffers from two conceptual issues. It does not solve the problem of the limited set of influential bits for every BR PUF \cite{Ganjietal_2016}. The relationship between the final and original challenges is not complex enough because their hamming distance will be constant all the time. Therefore, it is expected that DL modeling attacks can overcome this obfuscation and learn the appropriate transformation of input features as will be shown in the results section.

\subsection{Obfuscated PUF architecture 2 (N-to-1 Shuffled-Challenge Hierarchical XOR BR PUF)}
\label{sec:OBF-ARCH2}
\begin{figure}[ht]
	\centering
	\includegraphics[trim=0cm 0cm 0cm 0cm, clip=true,width=0.8\linewidth]{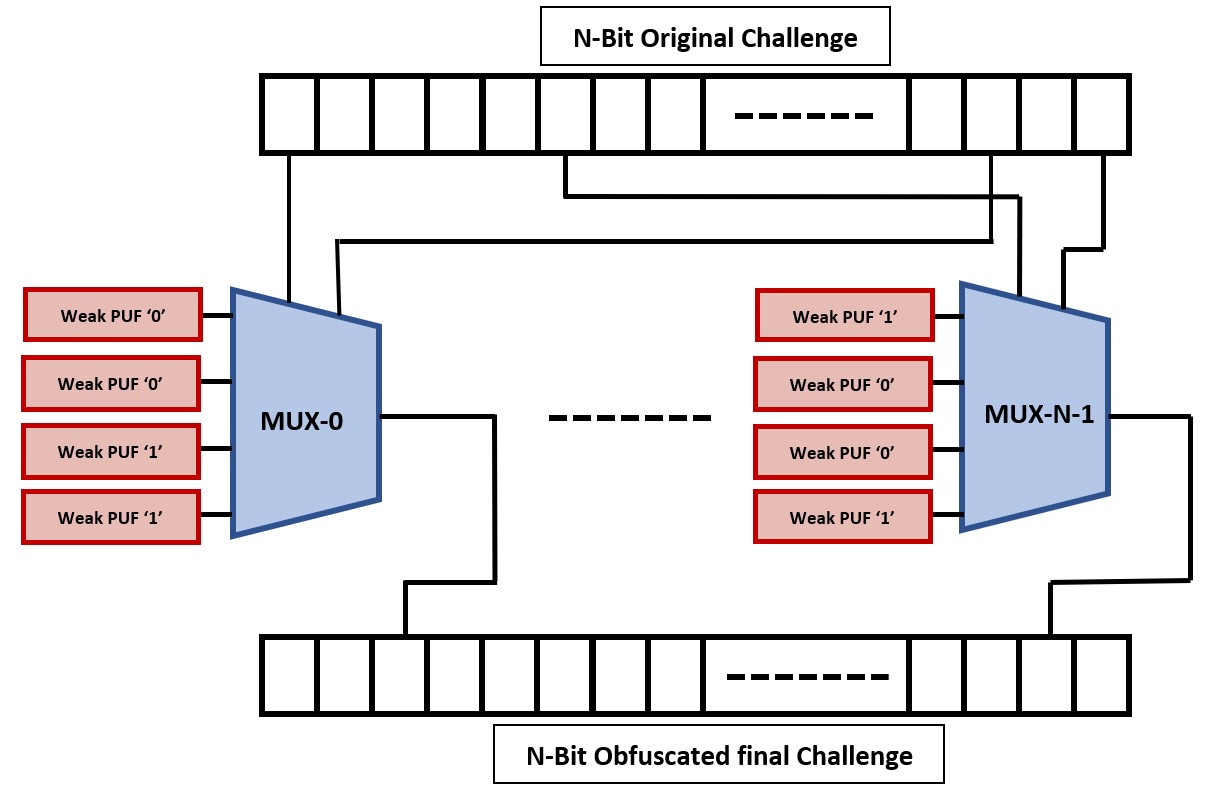}
	\caption{The 2-to-1 Shuffled-Challenge Hierarchical XOR BR PUF Architecture}
	\label{fig:fig_SCHPUF}
\end{figure}

In this obfuscation technique, we try to improve the architecture in terms of the limited number of influential bits and the relationship between the original and final challenges. Firstly the N-to-1 term denotes that every N bits of the original challenge are responsible for determining the value of one final challenge bit. Fig~\ref{fig:fig_SCHPUF} shows an example of a 2-to-1 Shuffled-Challenge Hierarchical XOR BR PUF, where every final challenge bit is derived by a multiplexer which its select inputs are a pair of the original challenge bits selected randomly. The conditions to be met are that no pairs are repeated and that every bit is involved in determining the value of two different final challenge bits. For this architecture, there is no single bit responsible for determining the value of any specific PUF stage. Furthermore, every bit impacts two different bits of the final challenge. Hence, the number of influential bits should increase and their relationship is more complex. Additionally, In order to increase the randomness between the original and final challenges, the position of every final challenge bit at which a pair of bits determine its value is randomly assigned for every chip. Moreover, the four memory-PUF connected to every multiplexer inputs must represent 50\% '1' values and their ordering is randomly selected from the six available choices that have two '1's and '0's. These design rules allowed this architecture to show a significant resistance against DL modeling attacks as will be shown in the results section.  

\section{Deep Learning Network Architecture and Experimental Setup}\label{DNN-Arch}

\subsection{Deep neural network architecture}
\begin{figure}[ht]
	\centering
	\includegraphics[trim=0cm 0cm 0cm 0cm, clip=true,width=0.8\linewidth]{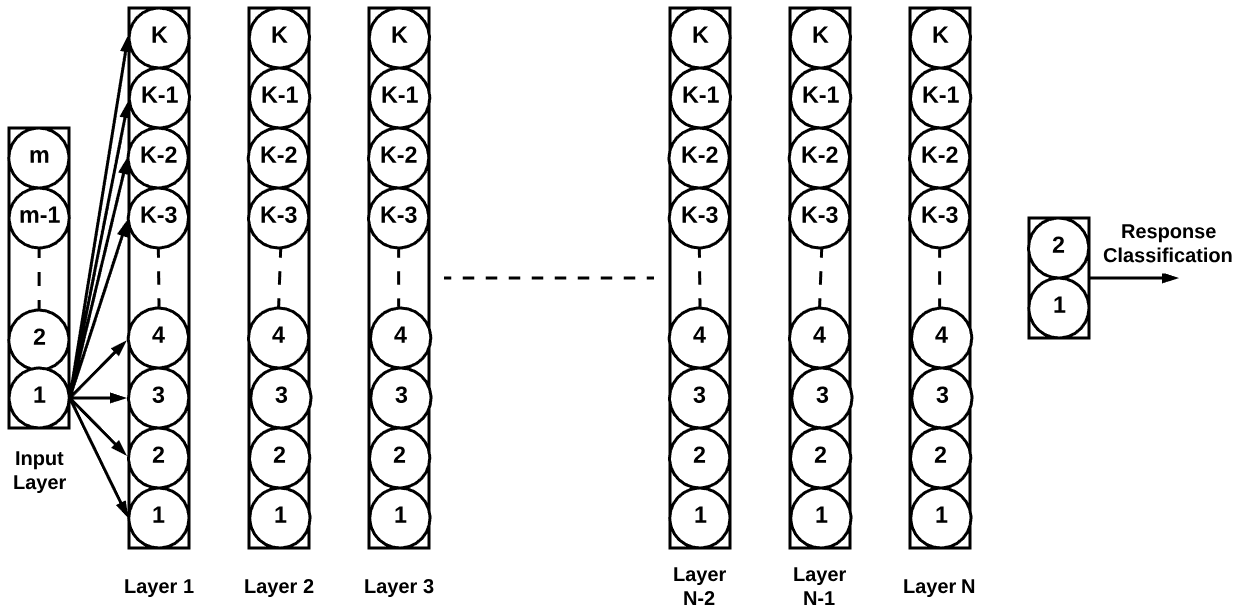}
	\caption{The DNN network architecture used in modeling attacks. All layers are fully connected layers.}
	\label{fig:fig10}
\end{figure}

Deep neural networks (DNNs) are artificial networks with multiple hidden layers between the input and output that can model complex non-linear relationships among inputs better than shallow ANNs. These types of networks start to build up a complete inference about the complex problem by gaining partial knowledge through the multiple hidden layers and aggregate them together at the end to provide an accurate classification/decision. For example, when a specific network tries to classify an object in image processing applications, its shallow layers extract features of edges and contours. Then, the deeper layers connect between features to construct shapes and classify the object at the output layer. Hence, the problem of modeling a complex PUF architecture can be solved using this approach. The DNN can learn the complex relationships among different stages by discovering the easier correlations between challenge features first and build upon that through the network to classify the final PUF response. The type of DNN used in this work is the feed-forward network, which means the flow of data goes in one direction through multiple layers from input to output as shown in Figure~\ref{fig:fig10}. 

Figure~\ref{fig:fig10} shows the deep neural network architecture used in the modeling attacks against XOR BR PUF and XOR TBR PUF. All layers of the network are fully connected, which means that every hidden neuron in layer n is connected to all neurons in layer n+1. Every connection represents a weight that reflects the neuron effect from the preceding layer on the output produced by the neuron in the next layer. In the graph, 'm' represents the number of input features which is one of the values 64, 128, and 256 depending on the PUF number of stages. The number of hidden neurons in every layer is represented by 'K' and the network depth is denoted by 'N', which corresponds to the number of fully connected layers in the network. Although the values of K and N were varied in the experiments to study the DNN scalability, the network used to report the modeling accuracies results in section 5, has N = 12 and K = 2000. Finally, a dropout layer was placed between the last fully connected layer and the output to help the network generalization and avoid the problem of over-fitting.

In DNN, convolutional layers are used for extracting desired features from input data. It applies a convolution process using a set of filters on input data to detect important features related to the task.
Hence, the shallow layers are usually convolutional layers to extract primitive features and reduce input size then deeper layers are fully connected. This approach works in object recognition and classification tasks due two main reasons. Firstly, features in input images have a locality property, therefore, it is more efficient to apply smaller feature sized windows to find pixels correlations. Secondly, the convolutional layers are less computation-intensive. However, the locality property does not exist in the context of modeling complex PUF architectures. There might be correlations between different PUF stages with distant positions or even among stages from different PUFs (In case of having XOR PUF architecture). Furthermore, these correlations will change from one PUF instance to another, which makes using convolutional layers not practical. Hence, using fully connected layers in the network is like the brute-force technique to extract features that help in modeling PUF response.

\subsection{DNN training parameters}
In this subsection we discuss the DNN parameters used in our training processes. Firstly we used the logistic function in our classification layer. It outputs a probability distribution of the problem binary classes, which in our case a two-class PUF response. The logistic function $(\sigma)$ defined in equation~\ref{eq:eq_logistic} maps the input x to an output between 0 and 1. This value represents the probability that problem output will belong to a specific class given the input x value as shown in equation~\ref{eq:eq_prob_logistic}. The cross entropy is used as the loss function, which measures the likelihood of a given set of parameters $\theta$ of the model can result in a prediction of the correct class of each input sample. The network tries to maximize this likelihood function by using adaptive moment estimation (Adam) optimization algorithm proposed by Kingma, D. and Ba, J.  \cite{Kingma2014_date}. Experiments showed that it converges faster and introduces better accuracy results than the normal gradient decent algorithm. The learning rate is usually between 0.0001 and 0.00001 for the training accuracy to be stable. For every run we do 1M iterations and a stopping condition if training accuracy is not changing to the 4th fractional digit for 5 consecutive times. Furthermore, if there is no convergence after 1M iterations, we re-run it again, but for almost every training process convergence would occur within the first 100K iterations. We evaluate our models using the accuracy metrics, which measures how many correct prediction cases out of the whole test set.  

\begin{equation}
\label{eq:eq_logistic}
\sigma(x) = \frac{1}{1 + e^{-x}}
\end{equation}

\begin{equation}
\label{eq:eq_prob_logistic}
\begin{split}
P(t = 1|x) = \sigma(x) = \frac{1}{1 + e^{-x}} , \\ P(t = 0|x) = 1 - \sigma(x) = \frac{1}{1 + e^{-x}}
\end{split}
\end{equation}

\subsection{Modeling attacks using SVM with polynomial kernel}
Experiments include attacks using SVM to test the resistance of implemented XOR BR and XOR TBR PUFs against conventional ML attacks and to further justify the need for using DL techniques. A script is written using python and scikit-learn library to provide SVM modeling functionality. Note that a grid search was conducted first to tune the SVM model parameters. Moreover, a polynomial kernel with degree four (equivalent to the number of XOR inputs) is used as was done by Xu et al in \cite{Xu_2015}. 
\subsection{Hardware and software experimental setup}
Experiments are conducted using three Mojo V3 boards, each containing a Spartan 6 FPGA \cite{Spartan_6} (45nm process technology). All CRPs are generated using an LFSR as mentioned in Section~\ref{sec:Hardware}. SVM attacks were executed on Intel 8th Gen I7-8250 CPU with 16GB RAM of memory, and DL attacks were executed using Nvidia GeForce GTX 1080 Ti GPU card with 11GB RAM.The Tensorflow platform was used to develop the network architectures, training, and evaluation tasks. Finally, scikit-learn library was used to implement SVM.

\section{Results and Discussion}\label{results}

\subsection{DL Modeling attack results}\label{DL-results}
As mentioned earlier, modeling attacks were launched on 24 instances of 4-input XOR BR \& TBR PUF with varying stage size (64, 128, 256) and 64-stage obfuscated XOR BR PUFs, which all were implemented on three mojo V3 boards. The DNN used for the attack had 12 fully connected layers and 2000 hidden neurons in every layer. Training set size was varied between 5K to 1M CRPs. Furthermore, The same PUF instances were attacked using SVM with a polynomial kernel of degree four to account for the 4-input XOR function \cite{Xu_2015}. Finally, the test set size was 100K CRPs for all experiments and CRPs were randomly generated as discussed in Section~\ref{sec:Hardware}. Table~\ref{tab:BR-results} shows the modeling accuracy results for all XOR BR PUF instances implemented on chip-1 using both approaches (DL, SVM). Note that similar results were obtained for the instances implemented on Chips 2 and 3.  
\begin{table}[ht]
	\caption{4-input XOR BR PUF modeling accuracy}
	\label{tab:BR-results}
\begin{center}
\begin{tabular}[ht]{c|c|c|c|c|c|c}
			\hline 
			\rule[-1ex]{0pt}{2.5ex}  \multirow{3}{*}{Train. Size} & \multicolumn{6}{c}{PUF Size \& Modeling Technique}  \\ 
			\cline{2-7}
			\rule[-1ex]{0pt}{2.5ex}  & \multicolumn{2}{c|}{64-Bit} &  \multicolumn{2}{c|}{128-Bit}  &  \multicolumn{2}{c}{256-Bit}  \\ 
			\rule[-1ex]{0pt}{2.5ex}  & \multicolumn{1}{c}{DL} & \multicolumn{1}{c|}{SVM} & \multicolumn{1}{c}{DL} & \multicolumn{1}{c|}{SVM} & \multicolumn{1}{c}{DL} & \multicolumn{1}{c}{SVM} \\ 
	\hline 
	\rule[-1ex]{0pt}{2.5ex}5K   & 67.6\% & 53.2\% & 51.8\% & 51.1\% & 88.8\% & 53.3\% \\ 
	\hline 
	\rule[-1ex]{0pt}{2.5ex}10K  & 98.1\% & 54.1\% & 85.9\% & 51.8\% & 94.6\% & 53.3\% \\ 
	\hline 
	\rule[-1ex]{0pt}{2.5ex}20K  & 99.2\% & 54.6\% & 96.7\% & 52.4\% & 96.4\% & 53.3\% \\ 
	\hline 
	\rule[-1ex]{0pt}{2.5ex}50K  & 99.1\% & 55.2\% & 98.2\% & 53.7\%	& 96.4\% & 53.4\% \\ 
	\hline 
	\rule[-1ex]{0pt}{2.5ex}100K & 98.1\% & 56.3\% & 98.8\% & 54.3\% & 96.7\% & 53.4\% \\ 
	\hline 
	\rule[-1ex]{0pt}{2.5ex}1M   & 99.5\% & N/A & 99.1\% & N/A & 99.3\% & N/A \\ 
	\hline 
\end{tabular} 
\end{center}
\end{table}

It is shown in the table that DL modeling was successful in breaking the security of all XOR BR PUF instances. Furthermore, training using 100K CRPs was enough to reach the accuracy boundary of 99\% in the case of 64 stages and 1M CRPs to reach this accuracy value for 128 and 256 stages. On the other hand, the SVM technique failed to successfully model the PUFs and its accuracy matched what was reported in previous literature for the same training size \cite{Xu_2015}. The maximum modeling accuracy that could be reached by SVM was 62.4\% for the 64 stages PUF implemented on chip-2 using 100K CRPs for training. Moreover, SVM performance got worse when attempting to model PUFs with bigger stage sizes, while DNN did not seem to be affected by that. Instead, The DL modeling technique was powerful enough to get an accuracy $>$ 95\% using 20K CRPs for training. This is relatively a small dataset size and generally, deep networks need more samples to train on. In contrast, results show that SVM failed to get higher than the 62\% accuracy using up to 100K CRPs.

Table~\ref{tab:TBR-results} shows the modeling attacks results against 4-input XOR TBR PUF on Chip-1 with varying stage size (64, 128, 256). Similarly, DL modeling techniques could break the security of the XOR TBR PUF, while SVM showed nearly the same performance as in XOR BR PUFs. However, the models could achieve accuracy $>$ 95\% using more CRPs for training than XOR BR PUFs (50K CRPs). Furthermore, the maximum accuracy achieved was slightly less than or equal to 99\%. This is expected because TBR PUFs involve all inverters in its operation, hence a slightly more complicated architecture than BR PUFs with the same number of challenge bits. In addition, all instances implemented on chips 2 and 3 showed similar behavior as the PUFs realized on chip 1.

\begin{table}[ht]
	\caption{4-input XOR TBR PUF modeling accuracy}
	\label{tab:TBR-results}
	\begin{center}
		\begin{tabular}[ht]{c|c|c|c|c|c|c}
			\hline 
			\rule[-1ex]{0pt}{2.5ex}  \multirow{3}{*}{Train. Size} & \multicolumn{6}{c}{PUF Size \& Modeling Technique}  \\ 
			\cline{2-7}
			\rule[-1ex]{0pt}{2.5ex}  & \multicolumn{2}{c|}{64-Bit} &  \multicolumn{2}{c|}{128-Bit}  &  \multicolumn{2}{c}{256-Bit}  \\ 
			\rule[-1ex]{0pt}{2.5ex}  & \multicolumn{1}{c}{DL} & \multicolumn{1}{c|}{SVM} & \multicolumn{1}{c}{DL} & \multicolumn{1}{c|}{SVM} & \multicolumn{1}{c}{DL} & \multicolumn{1}{c}{SVM} \\ 
			\hline 
			\rule[-1ex]{0pt}{2.5ex}5K   & 84.4\% & 56.7\% & 63.3\% & 53.7\% & 71.5\% & 59.4\% \\ 
			\hline 
			\rule[-1ex]{0pt}{2.5ex}10K  & 85.9\% & 58.5\% & 76.4\% & 55.3\% & 87.2\% & 60.9\% \\ 
			\hline 
			\rule[-1ex]{0pt}{2.5ex}20K  & 94.2\% & 59.8\% & 85.7\% & 57.7\% & 94.9\% & 62.4\% \\ 
			\hline 
			\rule[-1ex]{0pt}{2.5ex}50K  & 97.1\% & 60.6\% & 95.6\% & 59.4\%	& 97.1\% & 63.5\% \\ 
			\hline 
			\rule[-1ex]{0pt}{2.5ex}100K & 97.3\% & 60\% & 96.7\% & 62.5\% & 97.7\% & 62.9\% \\ 
			\hline 
			\rule[-1ex]{0pt}{2.5ex}1M   & 98.8\% & N/A & 98.8\% & N/A & 98.7\% & N/A \\ 
			\hline 
		\end{tabular} 
	\end{center}
\end{table}

\begin{table}[ht]
	\caption{DL modeling accuracy(\%) of obfuscated 64-Bit 4-input XOR BR PUF architectures on Chip-1.}
	\label{tab:OBF-results}
	\begin{center}
		\begin{tabular}[ht]{c|c|c|c}
			\hline 
		 	Training size & 20K & 50K & 100K \\ 
			\hline 
			Obfuscated Architecture 1 & 97.1\% & 98\% & 99.1\% \\ 
			\hline 
			Obfuscated Architecture 2 & 58.2\% & 76.5\% & 82.3\% \\ 
			\hline 
		\end{tabular}
	\end{center} 
\end{table}
 
Table~\ref{tab:OBF-results} shows the modeling attacks results against obfuscated 4-input XOR BR PUF architectures mentioned in Section~\ref{sec:obfuscated} with stage size = 64. Similar to non-obfuscated architectures, DL attacks could break the obfuscated PUF Architecture 1 (Hierarchical XOR BR PUF). DL networks could learn the hidden relationship between the original and final challenges because the number of inverted bits and their positions is always constant. Hence, It is easy for the DNN to realize the transformation of input features. On the other hand, Obfuscated PUF Architecture 2 (2-to-1 Shuffled-Challenge Hierarchical XOR BR PUF) showed significant resistance against DL attacks. Reducing accuracy by nearly 40\% using 20K training CRPs and maximum accuracy achieved is 82\% using 100K CRPs. Furthermore, it was noted that training error was small in all cases but test error of the trained model was worse, which means that the obtained model generalization gap was increased. This shows that the obfuscation technique used in subsection~\ref{sec:OBF-ARCH2} was relatively successful in increasing the randomness between the original and final challenges and reducing the effect of influential bits. Further analysis is required to enhance the resistance of this architecture against DL attacks and study the hardware and power overhead. However, it is worth mentioning that the multiplexers used to modify the challenge can be used for all PUF instances on the same chip, which reduces their overhead in terms of both hardware and power. Finally, note that as previous cases, instances implemented on chip2 and 3 showed similar performance.
\subsection{DNN scalability Vs. increasing PUF stages}\label{scalability results}
The DNN architecture is built to mimic the way PUFs operate to facilitate the learning process therefore, the hidden neurons in one layer can be considered as the stages of one PUF. Similarly, the number of layers can be used to reflect the number of XOR inputs. Consequently, a question arises about the network parameters and whether they should be the same as the number of PUF stages and XOR inputs or not. Note that increasing the network size can enhance or complicate the task for the network to learn the non-linear relationships among all stages. Accordingly, more experiments were executed to study the DNN parameters scalability (i.e. number of layers and number of hidden neurons per layer) with respect to the PUFs complexity and number of stages. Hence, the same modeling attack was invoked while varying the DNN number of layers (1,4,8,12) and the hidden neurons per layer (64, 128, 256, 512, 1024, 2048). Then, observe what the maximum accuracy will be and how long the network will take to converge. The training set size was chosen to be 1M CRPs in order to guarantee that obtained results depend on how the network architecture is varied not because of not enough training samples to train on. Total 432 training processes were run on all the 4-input XOR BR and XOR TBR PUF instances and Table~\ref{tab:Scal-results} shows the network configurations that achieved maximum accuracy with minimum train time for every type.  

\begin{table}[ht]
	\caption{DNN Scalability Analysis On 4-input XOR BR and TBR PUFs. Showing network configuration that achieved best accuracy and minimum training time}
	\label{tab:Scal-results}
	\begin{center}
		\begin{tabular}[ht]{C{1.8cm}|C{1.2cm}|C{1.2cm}|C{1.2cm}|C{1.2cm}}
			\hline 
			\rule[-1ex]{0pt}{2.5ex}PUF Type  & Accuracy & Layers & hidden neurons & train. time (min)  \\ 
			\hline 
			\rule[-1ex]{0pt}{2.5ex}64-XOR-BR  & 99.5\% & 8 & 1024 & 4.8 \\ 
			\hline 
			\rule[-1ex]{0pt}{2.5ex}128-XOR-BR   & 99.2\% & 8 & 1024 & 20 \\ 
			\hline 
			\rule[-1ex]{0pt}{2.5ex}256-XOR-BR   & 99.1\% & 12 & 2048 & 15 \\ 
			\hline 
			\rule[-1ex]{0pt}{2.5ex}64-XOR-TBR  & 98.8\% & 8 & 1024 & 6.8 \\ 
			\hline 
			\rule[-1ex]{0pt}{2.5ex}128-XOR-TBR  & 98.8 & 8 & 1024 &  7.6\\ 
			\hline 
			\rule[-1ex]{0pt}{2.5ex}256-XOR-TBR  & 99\% & 4 & 2048 & 9.6\\ 
			\hline 
		\end{tabular} 
	\end{center}
\end{table}

Obtained results show many interesting findings. Firstly, single layer NNs failed to model any instance successfully and modeling accuracy ranged between 55\% - 62\% except for one 256 XOR BR PUF instance that achieved 70\% accuracy. This confirms the results obtained from LDA analysis discussed in Subsection~\ref{Motive}, which showed that response classes are linearly inseparable for these PUF types. Furthermore, Table~\ref{tab:Scal-results} shows that for most cases maximum accuracy could be achieved using smaller networks than the one used in our initial experiments. This means the same results could have been achieved in less training time than the bigger DNN. 

 Moreover, obtained results showed that network configurations with 4 layers and 2048 hidden neurons and 8 layers with any hidden size can achieve modeling accuracy $>$ 90\% for all PUF sizes. Note that there is a trade-off between convergence (number of iterations) and time, therefore for the 256 XOR TBR case in Table~\ref{tab:Scal-results} a network with 4 layers converged slower than a similar one with 12 layers but it still could finish faster because it has way less number of computations. Hence, given a constant number of XOR inputs, the network scales linearly in terms of layers and hidden neurons with respect to XOR BR PUF stage size.

As an illustration, the number of weights to be updated in the network architecture shown in Figure~\ref{fig:fig10} can be calculated using equation~\ref{eq:DL_param}. 

\begin{equation}\label{eq:DL_param}
\textrm{$\#$ of Weights = } (m \times K ) + ((N-1) \times K^{2}) + (K \times 2)
\end{equation}
\hspace{1cm}

Note that 'm' is the number of PUF stages, 'K' is the hidden neurons per layer, and 'N' is the number of fully connected layers. Therefore, adding a new layer has $K^{2}$ effect on the number of weights and consequently the computations. Hence, one should think of a balanced approach when constructing the DNN network for similar PUF architectures by giving more priority to increase neurons per layer first. The following step is to slightly increase the number of layers to achieve the best accuracy/time trade-off.

It is also worth mentioning that the training time shown in Table~\ref{tab:Scal-results} is for achieving the maximum accuracy. Using smaller network configurations could achieve a reasonable accuracy of 95\% in much less training time. Hence, these timing figures are limited by the hardware used and the acceptable accuracy desired.

\subsection{Discussion on successful DL attacks and countermeasures}\label{sec:DL_Discuss}
In order to understand why DL networks could model these PUF architectures while SVM and single layer NN failed, the sources of modeling errors should be identified. Hence, let $E(f)$ and $E_{n}(f)$ be the test error and the training error for any classifier f respectively. Furthermore, let F be the space of functions that can be expressed by deep neural networks, $f^{*}_{F}$ is the best classifier in the F space and $f^{*}$ is the best possible classifier. If $\hat f$ is the classifier function returned by the training algorithm, then its excess error from the best possible classifier $\varepsilon \triangleq E(\hat f) - E(f^{*})$ can be attributed to two main terms as shown in equation~\ref{eq:DL_eq} \cite{Fan_19}. 

\begin{equation}
\label{eq:DL_eq}
\varepsilon = [E(f^{*}_{F}) - E(f^{*})] + [E(\hat f_{n}) - E(f^{*}_{F})]
\end{equation}
\hspace{1cm}

The first term is called the approximation error and measures how well the desired function can be approximated by a neural network using training samples. DNNs decrease the approximation error because they can express the composition of nonlinear functions effectively through their stacked layers (near zero training error. $\sim$ 0.0001 in our case). It was shown in \cite{Rolnick_17} that deep networks have a linear relationship with the input data dimension with respect to hidden neurons per layer, while shallow networks require an exponential number of neurons. This, in fact, was confirmed in our scalability analysis, where networks with a deeper number of layers could reach the 99\% accuracy using neurons in the range of $\mathcal{O}(n)$ with respect to n-stage PUFs. The second term in equation~\ref{eq:DL_eq} refers to the estimation error, which measures how well the trained model performs on out of sample data (generalization capability). DNNs with a large number of parameters and fully connected layers can control the generalization gap (small test error, $<$1\% in our case), if the complexity of all functions in F space is not large \cite{Fan_19}.

As a result, the successful DL attacks against the BR PUF family can be attributed to several reasons. Firstly the effect of influential bits that was reported in \cite{Ganjietal_2016}, which decreased the architecture complexity. Hence, introducing the XOR relationship was not sufficient to increase the complexity and counter the DL attacks. Furthermore, despite the use of a simplified mathematical model that does not represent the PUF operation accurately, DNNs could overcome that because it can learn the appropriate transformation of input features to correctly predict the target. Moreover, the lack of randomness when attempting to hide the challenge bits to counter DNNs modeling capabilities. Consequently, the obfuscated architecture-1 could be attacked and modeled with accuracy similar to non-obfuscated ones, while obfuscated architecture-2 showed a far better resistance against the DL attacks because it was built to minimize the effects of the above-mentioned reasons. This could be achieved by allowing more than one bit from the original challenge to determine the value of every bit of the final input challenge supplied to the PUF. Additionally, every original challenge bit affects two different positions of the final input challenge. These modifications resulted in increasing the randomness between the original and final challenges and the architecture complexity due to spreading the influence of every challenge bit on more than one position. Hence, as was mentioned in the results section, the generalization gap was increased and the test accuracy of DL attacks was worse despite the low training error achieved. Further analysis is needed to develop more complex versions of this obfuscated PUF and statistical metrics that measure the desired complexity of architectures to increase the generalization gap and counter these types of DL attacks. 
\subsection{{The practicality of the DL attacks and applications}}\label{sec:DL_Practicality}
The discussion of DL modeling attacks involves the access to PUF, the number of CRPs required for a successful attack, and the power needed to read CRPs and perform the DL training and build a model. In the context of strong PUFs(i.e. BR PUF family), they are usually not protected against the process of sending out challenges and reading out the response to collect the CRPs necessary for the attack \cite{Ruhrmair_2013}\cite{Kulseng_2010}\cite{Akgun_2015}\cite{Xu_2018}. However, there have been other authentication protocols that hide the response using hash functions or other cryptographic schemes \cite{Gope_2018}\cite{Zhu_2019}. Therefore, for the DL modeling attacks to work successfully against these types of controlled PUF environment, The assumption is that the attacker gains physical access to the PUF. Furthermore, the response hiding technique may be overcome by probing the digital signals coming out of the PUF before being input to the cryptographic logic used to hide the response \cite{Ruhrmair_2010} and \cite{Ruhrmair_2013}. 

The results showed that an accuracy of 95\% could be achieved using 20K and 50K CRPs for XOR BR PUFs and XOR TBR PUFs respectively. This number is surprisingly small given the network architecture and large parameters used for training. However, it is comparable to the number of CRPs used in modeling attacks against 4-input XOR arbiter PUFs, which used  12K and 20K CRPs to break the 64 and 128-bit PUFs respectively\cite{Ruhrmair_2013}. Additionally, it is normal that electrical strong PUFs operate at frequencies of a few MHZ\cite{Ruhrmair_2010}. For example, the mojo FPGA chips operate at 50 MHZ and the maximum number of cycles needed for evaluation is 19K as mentioned in Table~\ref{tab:results}. Hence, with eased conditions and assuming it takes 20K cycle to read a response, it takes $\sim$ 7 mins to read 1M CRPs.

The task of collecting CRPs is not computationally intensive, therefore, reading CRPs from chips with limited hardware resources as in \cite{Zhu_2019} is applicable under the above-mentioned assumption. Additionally, as far as we know, there was not any power analysis for CRPs collection task in the previously published attacks. Although, the DNN training task is computationally intensive it is possible now to execute training tasks with minimized time and power cost using GPUs \& ASIC chips (e.g. Tensor processing unit TPU and other accelerators). For cases where PUF attacks are not possible ( the CRPs are hidden, silicon probing is not possible, or battery-power limits the number of CRPs), DNNs may be useful to measure post-silicon PUF security validation before the chip is employed in the field.

Other DL attacks were reported against variants of APUFs\cite{Khalafalla_2019}, which means that DL modeling attacks can be considered as a powerful tool in breaking the security of strong PUFs (i.e. APUF variants and XOR BR PUF family including obfuscated versions) using a simplified mathematical model. Furthermore, these attacks are practical with respect to the number of CRPs and the power needed to execute the attack. It is applicable to break wide range of security protocols that use strong PUFs for authentication \cite{Akgun_2015}\cite{Xu_2018}\cite{Zhu_2019}, key establishment\cite{Tuyls_2007}, and Oblivious transfer protocols\cite{Ruhrmair_2010_2}.

\section{Conclusion and Future Work}\label{conclusion}
In this research, the deep learning modeling technique was introduced as a powerful tool to attack and break the security of complex strong PUF architectures implemented on real FPGAs(24 instances on three chips). It was shown that DNN can be used along with a simplified mathematical model to attack 4-input XOR BR PUF, 4-input XOR TBR PUF with varying number of stages (64, 128, 256), and a 64-stage obfuscated Hierarchical XOR BR PUF  and could successfully model all their responses with modeling accuracy $\sim$ 99\%. The DL attacks on the XOR BR family are practical and easy to launch because of the hardware and software support that enables training tasks in a matter of minutes as discussed in subsection~\ref{sec:DL_Practicality}. Furthermore, all successfully attacked architectures needed only between 20K to 50K CRPs for training to achieve modeling accuracy $>$ 95\%. 

Even with the lack of an accurate mathematical model, DNNs were shown to provide better performance than SVM with polynomial kernel and single layer NNs likely due to the ability of the nonlinear stacked layers to learn the appropriate data transformations needed (as discussed at subsection~\ref{sec:DL_Discuss}). Additionally, a detailed analysis was conducted to study the scalability of DNNs used to model XOR BR and TBR PUF architectures. This analysis included 432 modeling attacks on all PUF instances using 24 network configurations for every instance. Results showed that maximum accuracy can be achieved using smaller network architectures and the number of hidden neurons per layer scale linearly with the increase of PUFs stage size (given that the number of XOR inputs is constant), which agrees with the approximation theory of DNN as discussed in subsection~\ref{sec:DL_Discuss}. 

The 2-to-1 Shuffled-Challenge Hierarchical XOR BR PUF was introduced as a new architecture to countermeasure the DL attacks by overcoming the inherent problem of influential bits in BR PUFs and increasing the randomness between the original and final challenge bits. Hence, increasing the architecture complexity and the generalization gap of the DL model. Attacks on 64-stage instances showed significance resistance and a promising first step towards an ideal resilience against DL attacks.

Future work will investigate the scaling of DNNs while varying the XOR input number. Moreover, the feasibility of DL attacks against new PUF architectures with different obfuscation techniques and countermeasures. Furthermore, more analysis is required to develop new statistical metrics to measure the appropriate level of obfuscation needed to consider a specific PUF architecture secure.

This research was supported in part by grants from NSERC.

\bibliographystyle{IEEEtran}
\bibliography{Host_2020_V2.0}

\begin{thebibliography}{10}
\providecommand{\url}[1]{#1}
\csname url@samestyle\endcsname
\providecommand{\newblock}{\relax}
\providecommand{\bibinfo}[2]{#2}
\providecommand{\BIBentrySTDinterwordspacing}{\spaceskip=0pt\relax}
\providecommand{\BIBentryALTinterwordstretchfactor}{4}
\providecommand{\BIBentryALTinterwordspacing}{\spaceskip=\fontdimen2\font plus
\BIBentryALTinterwordstretchfactor\fontdimen3\font minus
  \fontdimen4\font\relax}
\providecommand{\BIBforeignlanguage}[2]{{%
\expandafter\ifx\csname l@#1\endcsname\relax
\typeout{** WARNING: IEEEtran.bst: No hyphenation pattern has been}%
\typeout{** loaded for the language `#1'. Using the pattern for}%
\typeout{** the default language instead.}%
\else
\language=\csname l@#1\endcsname
\fi
#2}}
\providecommand{\BIBdecl}{\relax}
\BIBdecl

\bibitem{Lee_2004}
J.~W. {Lee}, , B.~{Gassend}, G.~E. {Suh}, M.~{van Dijk}, and S.~{Devadas}, ``A
  technique to build a secret key in integrated circuits for identification and
  authentication applications,'' in \emph{2004 Symposium on VLSI Circuits.
  Digest of Technical Papers (IEEE Cat. No.04CH37525)}, June 2004, pp.
  176--179.

\bibitem{Lim_2004}
L.~D., ``Extracting secret keys from integrated circuits,'' Master's thesis,
  MIT, 2004.

\bibitem{Maes_2015}
R.~Maes, V.~{Van Der Leest}, E.~{Van Der Sluis}, and F.~Willems, ``{Secure key
  generation from biased PUFs},'' \emph{Lecture Notes in Computer Science
  (including subseries Lecture Notes in Artificial Intelligence and Lecture
  Notes in Bioinformatics)}, vol. 9293, pp. 517--534, 2015.

\bibitem{Bohm_2012}
C.~B{\"o}hm and M.~Hofer, \emph{Physical unclonable functions in theory and
  practice}.\hskip 1em plus 0.5em minus 0.4em\relax Springer Science \&
  Business Media, 2012.

\bibitem{Katzenbeisser_2012}
S.~Katzenbeisser, {\"U}.~Kocaba{\c s}, V.~Ro{\v z}i{\' c}, A.~R. Sadeghi,
  I.~Verbauwhede, and C.~Wachsmann, ``{PUFs: Myth, fact or busted? A security
  evaluation of Physically Unclonable Functions (PUFs) cast in silicon},''
  \emph{Lecture Notes in Computer Science (including subseries Lecture Notes in
  Artificial Intelligence and Lecture Notes in Bioinformatics)}, vol. 7428
  LNCS, pp. 283--301, 2012.

\bibitem{Ruhrmair_2010}
\BIBentryALTinterwordspacing
U.~R{\"{u}}hrmair, F.~Sehnke, J.~{S {\"{o}}lter}, G.~Dror, S.~Devadas, and
  J.~{\"{u}}. Schmidhuber, ``{Modeling attacks on physical unclonable
  functions},'' \emph{Proceedings of the 17th ACM conference on Computer and
  communications security - CCS '10}, p. 237, 2010. [Online]. Available:
  \url{http://dl.acm.org/citation.cfm?id=1866307.1866335}
\BIBentrySTDinterwordspacing

\bibitem{Ruhrmair_2013}
\BIBentryALTinterwordspacing
U.~Ruhrmair, J.~Solter, F.~Sehnke, X.~Xu, A.~Mahmoud, V.~Stoyanova, G.~Dror,
  J.~Schmidhuber, W.~Burleson, and S.~Devadas, ``{PUF} modeling attacks on
  simulated and silicon data,'' \emph{Trans. Info. For. Sec.}, vol.~8, no.~11,
  pp. 1876--1891, Nov. 2013. [Online]. Available:
  \url{http://dx.doi.org/10.1109/TIFS.2013.2279798}
\BIBentrySTDinterwordspacing

\bibitem{BR_intro}
Q.~Chen, G.~Csaba, P.~Lugli, U.~Schlichtmann, and U.~R{\"u}hrmair, ``The
  bistable ring puf: A new architecture for strong physical unclonable
  functions,'' in \emph{2011 IEEE International Symposium on Hardware-Oriented
  Security and Trust}.\hskip 1em plus 0.5em minus 0.4em\relax IEEE, 2011, pp.
  134--141.

\bibitem{Chen_2012}
Q.~{Chen}, G.~{Csaba}, P.~{Lugli}, U.~{Schlichtmann}, and U.~{Rührmair},
  ``Characterization of the bistable ring puf,'' in \emph{2012 Design,
  Automation Test in Europe Conference Exhibition (DATE)}, March 2012, pp.
  1459--1462.

\bibitem{TBR_intro}
D.~Schuster and R.~Hesselbarth, ``Evaluation of bistable ring pufs using single
  layer neural networks,'' in \emph{International Conference on Trust and
  Trustworthy Computing}.\hskip 1em plus 0.5em minus 0.4em\relax Springer,
  2014, pp. 101--109.

\bibitem{Xu_2015}
X.~Xu, U.~R{\"{u}}hrmair, D.~E. Holcomb, and W.~Burleson, ``{Security
  evaluation and enhancement of Bistable Ring PUFs},'' \emph{Lecture Notes in
  Computer Science (including subseries Lecture Notes in Artificial
  Intelligence and Lecture Notes in Bioinformatics)}, vol. 9440, pp. 3--16,
  2015.

\bibitem{Ganjietal_2016}
F.~Ganji, S.~Tajik, F.~Fäßler, and J.-P. Seifert, ``Strong machine learning
  attack against pufs with no mathematical model,'' Cryptology ePrint Archive,
  Report 2016/606, 2016, \url{http://eprint.iacr.org/2016/606}.

\bibitem{Ganji_2016_2}
F.~Ganji, S.~Tajik, and J.-P. Seifert, ``Why attackers win: On the learnability
  of xor arbiter pufs,'' in \emph{Trust and Trustworthy Computing}, M.~Conti,
  M.~Schunter, and I.~Askoxylakis, Eds.\hskip 1em plus 0.5em minus 0.4em\relax
  Cham: Springer International Publishing, 2015, pp. 22--39.

\bibitem{Hospodar_2012}
G.~Hospodar, R.~Maes, and I.~Verbauwhede, ``Machine learning attacks on 65nm
  arbiter {PUF}s: Accurate modeling poses strict bounds on usability,'' in
  \emph{2012 IEEE International Workshop on Information Forensics and Security
  (WIFS)}, Dec 2012, pp. 37--42.

\bibitem{Tobisch_2015}
J.~Tobisch and G.~T. Becker, ``On the scaling of machine learning attacks on
  {PUFs} with application to noise bifurcation,'' in \emph{Radio Frequency
  Identification}, S.~Mangard and P.~Schaumont, Eds.\hskip 1em plus 0.5em minus
  0.4em\relax Cham: Springer International Publishing, 2015, pp. 17--31.

\bibitem{Freund_1997}
\BIBentryALTinterwordspacing
Y.~Freund and R.~E. Schapire, ``A decision-theoretic generalization of on-line
  learning and an application to boosting,'' \emph{J. Comput. Syst. Sci.},
  vol.~55, no.~1, pp. 119--139, Aug. 1997. [Online]. Available:
  \url{http://dx.doi.org/10.1006/jcss.1997.1504}
\BIBentrySTDinterwordspacing

\bibitem{hastie_01}
T.~Hastie, R.~Tibshirani, and J.~Friedman, \emph{The Elements of Statistical
  Learning}, ser. Springer Series in Statistics.\hskip 1em plus 0.5em minus
  0.4em\relax New York, NY, USA: Springer New York Inc., 2001.

\bibitem{capacitance}
D.~Yamamoto, M.~Takenaka, K.~Sakiyama, and N.~Torii, ``Security evaluation of
  bistable ring pufs on fpgas using differential and linear analysis,'' in
  \emph{2014 Federated Conference on Computer Science and Information
  Systems}.\hskip 1em plus 0.5em minus 0.4em\relax IEEE, 2014, pp. 911--918.

\bibitem{metrics}
U.~Chatterjee, R.~S. Chakraborty, and D.~Mukhopadhyay, ``A puf-based secure
  communication protocol for iot,'' \emph{ACM Transactions on Embedded
  Computing Systems (TECS)}, vol.~16, no.~3, p.~67, 2017.

\bibitem{Ma_2018}
Q.~Ma, C.~Gu, N.~Hanley, C.~Wang, W.~Liu, and M.~O'Neill, ``{A machine learning
  attack resistant multi-PUF design on FPGA},'' \emph{Proceedings of the Asia
  and South Pacific Design Automation Conference, ASP-DAC}, vol. 2018-Janua,
  pp. 97--104, 2018.

\bibitem{Kingma2014_date}
D.~P. Kingma and J.~Ba, ``Adam: A method for stochastic optimization,''
  \emph{arXiv preprint arXiv:1412.6980}, 2014.

\bibitem{Spartan_6}
Xilinx, \emph{Spartan-6 {FPGA} Configurable Logic Block User Guide (UG384 )},
  Xilinx.

\bibitem{Fan_19}
J.~{Fan}, C.~{Ma}, and Y.~{Zhong}, ``{A Selective Overview of Deep Learning},''
  \emph{arXiv e-prints}, p. arXiv:1904.05526, Apr 2019.

\bibitem{Rolnick_17}
\BIBentryALTinterwordspacing
D.~Rolnick and M.~Tegmark, ``The power of deeper networks for expressing
  natural functions,'' \emph{CoRR}, vol. abs/1705.05502, 2017. [Online].
  Available: \url{http://arxiv.org/abs/1705.05502}
\BIBentrySTDinterwordspacing

\bibitem{Kulseng_2010}
L.~{Kulseng}, Z.~{Yu}, Y.~{Wei}, and Y.~{Guan}, ``Lightweight mutual
  authentication and ownership transfer for rfid systems,'' in \emph{2010
  Proceedings IEEE INFOCOM}, March 2010, pp. 1--5.

\bibitem{Akgun_2015}
\BIBentryALTinterwordspacing
M.~Akg\"{u}n and M.~U. \c{C}aglayan, ``Providing destructive privacy and
  scalability in rfid systems using pufs,'' \emph{Ad Hoc Netw.}, vol.~32,
  no.~C, pp. 32--42, Sep. 2015. [Online]. Available:
  \url{http://dx.doi.org/10.1016/j.adhoc.2015.02.001}
\BIBentrySTDinterwordspacing

\bibitem{Xu_2018}
\BIBentryALTinterwordspacing
H.~Xu, J.~Ding, P.~Li, F.~Zhu, and R.~Wang, ``A lightweight rfid mutual
  authentication protocol based on physical unclonable function,''
  \emph{Sensors}, vol.~18, no.~3, p. 760, Mar 2018. [Online]. Available:
  \url{http://dx.doi.org/10.3390/s18030760}
\BIBentrySTDinterwordspacing

\bibitem{Gope_2018}
P.~{Gope}, J.~{Lee}, and T.~Q.~S. {Quek}, ``Lightweight and practical anonymous
  authentication protocol for rfid systems using physically unclonable
  functions,'' \emph{IEEE Transactions on Information Forensics and Security},
  vol.~13, no.~11, pp. 2831--2843, Nov 2018.

\bibitem{Zhu_2019}
F.~Zhu, P.~Li, H.~Xu, and R.~Wang, ``A lightweight rfid mutual authentication
  protocol with puf,'' \emph{Sensors}, vol.~19, p. 2957, 07 2019.

\bibitem{Khalafalla_2019}
M.~{Khalafalla} and C.~{Gebotys}, ``Pufs deep attacks: Enhanced modeling
  attacks using deep learning techniques to break the security of double
  arbiter pufs,'' in \emph{2019 Design, Automation Test in Europe Conference
  Exhibition (DATE)}, March 2019, pp. 204--209.

\bibitem{Tuyls_2007}
P.~Tuyls and B.~{\v{S}}kori{\'{c}}, \emph{Strong Authentication with Physical
  Unclonable Functions}.\hskip 1em plus 0.5em minus 0.4em\relax Berlin,
  Heidelberg: Springer Berlin Heidelberg, 2007, pp. 133--148.

\bibitem{Ruhrmair_2010_2}
U.~R\"{u}hrmair, ``Oblivious transfer based on physical unclonable functions,''
  in \emph{Proceedings of the 3rd International Conference on Trust and
  Trustworthy Computing}, ser. TRUST'10.\hskip 1em plus 0.5em minus 0.4em\relax
  Berlin, Heidelberg: Springer-Verlag, 2010, pp. 430--440.

\end{thebibliography}

\end{document}